# Plasma stratification in AC discharges in noble gases at low currents

Vladimir I. Kolobov[1,2] and Robert R. Arslanbekov[2]

[1]*University of Alabama in Huntsville, Huntsville, AL, USA*
[2]*CFD Research Corporation, Huntsville, AL, USA*

## Abstract

A hybrid kinetic-fluid model is used to study plasma stratification in alternating current (AC) discharges in noble gases at low plasma densities. Self-consistent coupled solutions of a nonlocal kinetic equation for electrons, a drift-diffusion equation of ions, and a Poisson equation for the electric field are obtained for a positive column and the entire discharge with near-electrode sheaths. Standing striations are obtained for the reduced values of electric fields, $E/p$, corresponding to the inelastic energy balance of electrons in a range of driving frequencies. An analog of Novak's law, $\Lambda \approx \varepsilon_1/(e\langle E\rangle)$ (striation length $\Lambda$ proportional to the excitation threshold of atoms $\varepsilon_1$ and inversely proportional to the mean square root of the electric field $\langle E\rangle$, $e$ is the electron charge) is observed in simulations, indicating the nonlocal nature of standing striations in AC discharges at low plasma densities. Stratified plasma operates in a dynamic regime for various driving frequencies. In this regime, ions respond to the time-averaged electric field, whereas electrons react to the instantaneous electric field. The disparity of time scales between ambipolar diffusion, which occurs at the ion time scale, and electron kinetics in the coordinate-energy phase space, which occurs at the free electron diffusion time scale, produces complicated fluxes in the phase space (due to electron heating, energy loss in collisions, and ionization processes) that are responsible for the stratification. Our paper emphasizes the need for the kinetic approach to analyze stratification phenomena in AC discharge of noble gases. It promotes an efficient method for the kinetic treatment of electrons that is an alternative to the commonly used Particle-in-Cell (PIC) method.

## I. Introduction

Plasma stratification into bright and dark layers along electric current has been commonly observed in noble gases. In direct current (DC) discharges, striations usually move at high speeds and are not visible to the naked eye [1,2,3]. Standing striations have been observed in symmetric AC discharges; they move slowly in the presence of any asymmetry or a small DC component of the electric field. Striations have been experimentally observed in capacitively coupled plasma (CCP) [4,5,6], inductively coupled plasma (ICP) [7,8,9], and dielectric barrier discharges (DBD) [10,11] at various gas pressures and driving frequencies. A similar nature of moving striations in DC discharges and standing striations in high-frequency AC discharges has been demonstrated by Nedospasov [1,2] using linear analysis of a dispersion equation for strata in noble gases. However, plasma stratification remains a subject of active research. Recent experiments and PIC simulations have demonstrated discharge stratification in atomic [12,13] and molecular [14,15] gases. The interpretation of the observed phenomena in noble gases used an extended fluid model [16,17], which cannot be well justified for the reasons described below. Therefore, the nature of AC plasma stratification in noble gases has remained unclear.

Timofeev [18] first proposed a "thermocurrent" instability that could lead to the stratification of DC and AC discharges. Instability occurs when the thermo-diffusion electron flux is directed opposite to the diffusion flux and exceeds its magnitude. The review [19] and references cited therein explain that the maximal increment of this instability is observed for short wavelengths compared to the electron energy relaxation length. Under such conditions, the fluid description of electrons cannot be justified, and a kinetic model using the PIC method [20] or a solution of a nonlocal kinetic equation [21,22] should be used. When Coulomb collisions among electrons are essential, the fluid model for electrons can be well justified and used to explain plasma stratification at high plasma densities [23]. This model has been recently applied to analyze ICP stratification [12]. In the present paper, a hybrid kinetic-fluid model previously used for moving striations in DC discharges at low plasma densities [24] is adapted to simulations of standing striations in AC discharges.

We describe an alternative to the PIC method, a grid-based Fokker-Planck (FP) kinetic solver for treating nonlocal kinetic effects in phase space. The FP approach based on the Lorentz model (aka two-term spherical harmonics expansion) has been theoretically justified and used in numerous publications [25]. Frequent elastic collisions of electrons with atoms allow reducing the Boltzmann kinetic equation to the FP kinetic equation in the energy-coordinate phase space of lower dimensionality [26]. In our simulations, the FP kinetic solver for electrons is coupled to a drift-diffusion model for ions and a Poisson solver for the electric field. In plasma, electrons and ions move at equal rates due to the ambipolar electric field, which ensures quasineutrality. The mass difference of electrons and ions produces a disparity of scales for electrons' density and energy flow. Discrete energy loss of electrons for excitation of atoms plays a vital role in plasma stratification. The ionization of atoms by electron impact is slow because it must be balanced by (slow) ion loss. As a result, in spatially non-uniform plasma, electrons with different energies move independently in the phase space, creating nonlocal effects responsible for complicated dynamics of heat flow in plasmas. We demonstrate that the grid-based FP kinetic solver efficiently analyzes nonlocal kinetic effects in phase space.

We show that AC discharges in noble gases operate in a dynamic regime for a wide range of driving frequencies. The dynamic regime was previously analyzed for low-frequency ICP [27] and a striation-free AC positive column [28]. In this regime, ions respond to the time-averaged electric field, whereas electrons react to the instantaneous electric field. Electron kinetics in dynamic discharges is the most sophisticated yet not fully explored. Dynamic effects are associated with the disparity of time scales between ambipolar diffusion, which occurs at the ion time scale, and electron kinetics in phase space, which proceed at the *free* electron diffusion time scale. The spatially non-uniform plasma density profiles remain practically unchanged during the AC period in dynamic discharges. Still, complicated electron fluxes in the phase space occur due to electron heating by instantaneous AC electric fields, energy loss in collisions, and ionization processes. In the present paper, we consider the case of low ionization degree (plasma density) when the nonlinear processes associated with stepwise ionization, Coulomb collisions among electrons, and gas heating are insignificant.

The paper is organized as follows. Section II describes the computational model and boundary conditions. Section III describes the results for a positive column with periodic boundary conditions. Section IV analyses the impact of near-electrode sheaths on CCP stratification. Section V discusses the new findings and their importance. Section VI contains conclusions and identifies open questions and future research plans.

## II. Computational model

The hybrid model used in the present paper was first described in [29] and later adapted in [24] to study striations in DC discharges and the radial structure of positive columns in AC discharges [28]. This model is extended to analyze plasma stratification in AC discharges in noble gases.

### 1. Electron kinetics

Frequent elastic collisions of electrons with atoms in weakly ionized collisional plasma justify using the Lorentz model to reduce the Boltzmann kinetic equation to a Fokker-Planck kinetic equation in the energy-coordinate phase space. With the two-term spherical harmonics expansion, the Boltzmann kinetic equation for electrons can be reduced to a set of two coupled equations [30]:

$$\frac{\partial f_0}{\partial t} + \frac{v}{3}\nabla \cdot \boldsymbol{f}_1 - \frac{1}{3v^2}\frac{\partial}{\partial v}\left(v^2 \frac{e\boldsymbol{E}}{m} \cdot \boldsymbol{f}_1\right) = S_0 \quad (1)$$

$$\frac{\partial \boldsymbol{f}_1}{\partial t} + \nu \boldsymbol{f}_1 = -v\nabla f_0 + \frac{e\boldsymbol{E}}{m}\frac{\partial f_0}{\partial v} \quad (2)$$

Here, *e* and *m* are the unsigned electron charge and mass, **E** is the electric field vector, $\nu(v)$ is the transport collision frequency, and $S_0$ describes energy exchange in quasi-elastic and inelastic electron–atom collisions, electron-electron interactions, and ionization-recombination processes.

Equation (1) expresses the divergence of a flux $(\boldsymbol{\Phi}, \Gamma)$ in a 4D phase space $(\boldsymbol{r}, u)$:

$$\frac{\partial f_0}{\partial t} + \nabla \cdot \boldsymbol{\Phi} - \frac{1}{v}\frac{\partial}{\partial u}(v\Gamma) = S_0, \quad (3)$$

where $u = mv^2/(2e)$ is the volt-equivalent of the electron kinetic energy and

$$\boldsymbol{\Phi} = \frac{v}{3}\boldsymbol{f}_1 \quad , \quad \Gamma = \boldsymbol{E} \cdot \boldsymbol{\Phi} \quad (4)$$

The speed $v$ plays the role of the Lame coefficient. Equation (2) shows that $\boldsymbol{f}_1$ depends on the local value of the electric field, i.e., the two-term approximation results in Ohm's law [31]. The Electron Energy Probability Function (EEPF) $f_0$ is normalized on the electron density $n_e$ and has units eV$^{-3/2}$ cm$^{-3}$.

The general solution of Eq. (2) can be written in the form [32]:

$$\boldsymbol{f}_1(\boldsymbol{r}, v, t) = -ve^{-\nu t}\int_{-\infty}^{t} e^{\nu t'}\left[\nabla f_0 - \frac{e\boldsymbol{E}(\boldsymbol{r},t')}{mv}\frac{\partial f_0}{\partial v}\right]dt' \quad . \quad (5)$$

The main contribution to integral (5) comes from a small vicinity of the point $t$ about the size ~ $\nu^{-1}$. As the characteristic time scale for the time-variation $f_0$ is larger than $\nu^{-1}$, the derivatives can be removed from the integral sign in (5) to obtain:

$$\boldsymbol{f}_1(\boldsymbol{r}, v, t) = -\frac{v}{\nu}\left[\nabla f_0 - \mathcal{E}(t)\frac{\partial f_0}{\partial u}\right] \quad . \tag{6}$$

Here, we introduced an effective field:

$$\mathcal{E}(\boldsymbol{r}, t) = e^{-\nu t}\int_{-\infty}^{\nu t} e^{\tau}\boldsymbol{E}(\boldsymbol{r}, \tau)d\tau \quad , \tag{7}$$

which accounts for temporal dispersion. Using (6), the spatial flux becomes

$$\boldsymbol{\Phi} = -D_r\left(\nabla f_0 - \mathcal{E}\frac{\partial f_0}{\partial u}\right) \tag{8}$$

where $D_r = v^2/(3\nu)$ is the spatial diffusion coefficient in phase space. Substituting (4) into (1), we obtain a Fokker-Planck (FP) kinetic equation for $f_0(\boldsymbol{r}, u, t)$ in a 4D phase space $(\boldsymbol{r}, u)$:

$$\frac{\partial f_0}{\partial t} - \left(\nabla - \boldsymbol{E}\frac{1}{\sqrt{u}}\frac{\partial}{\partial u}\sqrt{u}\right)\cdot D_r\left(\nabla - \mathcal{E}\frac{\partial}{\partial u}\right) f_0 = S_0 \quad , \tag{9}$$

which can be rewritten in the form:

$$\frac{\partial f_0}{\partial t} - \nabla_4 \cdot (\mathbf{D}\nabla_4 f_0) = S_0 \quad , \tag{10}$$

where $\nabla_4$ denote the divergence and grad operators in the 4D phase space and $\mathbf{D}$ a diffusion tensor:

$$\mathbf{D} = D_r\begin{pmatrix} \mathbf{I}_3 & -\mathcal{E} \\ -\boldsymbol{E} & \mathcal{E}\cdot\boldsymbol{E} \end{pmatrix} . \tag{11}$$

Here $\mathbf{I}_3$ denotes the unit tensor in 3D configuration space. The kinetic equation (10) has already been used (in 2D phase space) to analyze the axial and radial structure of DC discharges and plasma stratification [28,24]. It also appeared in the study of fully ionized plasma [33].

The most favorable conditions for plasma stratification are observed when the two terms in the brackets of (9) are of the same order, which corresponds to a spatial scale $\Lambda = u/E$. In this case, electron fluxes in configuration space and energy described by the left part of Eq. (10) are of the same order. The time scale for $f_0(\boldsymbol{r}, u, t)$ variation is defined by a characteristic frequency $\nu_E = (E/u)^2 D_r$ due to electron diffusion in phase space and the frequency, $\nu_0 = S_0/f_0$ , which controls energy flow in collisions and electron generation processes. For slow variations of $E(t)$ compared to $\max\{1/\nu_E, 1/\nu_0\}$ , the time derivative in (9,10) can be neglected and $f_0(t)$ depends on the instantaneous value of the electric field. For fast variations of $E(t)$, $f_0$ depends on the average value of the electric field.

The right-hand side of Eqs. (90 and (10) describe the particle and energy flux due to collisions and electron generation processes. The electron energy loss in quasi-elastic collisions is defined as:

$$C_{el} = \frac{1}{\sqrt{u}}\frac{\partial}{\partial u}\left(u^{3/2}\delta\nu f_0\right) \quad (12)$$

where $\delta$ is the fraction of electron energy lost in quasi-elastic collisions. The excitation of an atomic state with the energy threshold $\varepsilon_1$ is described by

$$C_{ex} = -\nu^*(u)f_0(u) + \frac{\sqrt{u+\varepsilon_1}}{\sqrt{u}}\nu^*(u+\varepsilon_1)f_0(u+\varepsilon_1) \quad (13)$$

where $\nu^*(u)$ is the inelastic collision frequency. The operators (12) and (13) conserve the number of electrons. The total frequency of the EEPF relaxation in collisions is $\nu_u(u) = \delta\nu + \nu^*(u)$. For noble gases, $\nu_u(u)$ substantially depends on energy. At low energies, $u < \varepsilon_1$, this frequency is much lower than the elastic collision frequency $\nu_u(u) = \delta\nu \ll \nu$, where $\delta = 2m/M \ll 1$. At energies $u > \varepsilon_1$, this frequency is comparable to the frequency of inelastic collisions $\nu_u(u) \approx \nu^*(u)$. The latter is $\nu^*(u) \leq \nu(u)$, for the energy range of interest. For molecular gases, $\delta$ could be about the unity due to the excitation of rotational and vibrational states of molecules and $\nu_u = \delta\nu \approx \nu$ in the corresponding energy range.

The direct ionization by electron impact is described as:

$$C_{ion} = -\nu^{ion}(u)f_0(u) + 4\frac{\sqrt{2u+\varepsilon_{ion}}}{\sqrt{u}}\nu^{ion}(2u+\varepsilon_{ion})f_0(2u+\varepsilon_{ion}) \quad (14)$$

where $\nu^{ion}(u)$ is the ionization frequency and $\varepsilon_{ion}$ is the ionization threshold. The operator (14) describes the generation of new electrons, assuming that the kinetic energy is evenly distributed between the primary and secondary electrons after an ionization event. The frequency of ionization is the slowest process, as the production of electron-ion pairs in ionization events must be balanced by their loss, which is controlled by slow ion transport processes. For this reason, details of the energy redistribution between the primary and secondary electrons are not particularly important. Using Eq. (14), it is easy to switch off the electron generation but still consider the inelastic energy loss with the quantum of $\varepsilon_{ion}$. We have used this feature to analyze the effect of secondary electron generation on plasma stratification.

## 2. Discharge model

Below, we assume that plasma is contained in a long cylindrical tube of radius $R$ and is maintained by the axial electric field, $E(x,t)$. The charged particles are lost due to ambipolar diffusion and surface recombination at the wall. We rewrite the kinetic equation (10) for a 2D phase space $(x,u)$ as:

$$\frac{\partial f_0}{\partial t} - \nabla\cdot(\boldsymbol{D}_2\nabla f_0) = C_0 - C_{wall} \quad . \quad (15)$$

The diffusion tensor $\mathbf{D}_2$ is defined as:

$$\boldsymbol{D}_2 = D_r \begin{pmatrix} 1 & -\mathcal{E} \\ -E & \mathcal{E}E \end{pmatrix}, \tag{16}$$

and the radial loss $C_{wall}$ is added. The loss of electrons at the wall can be included in the form [34]:

$$C_{wall} = -\frac{1}{3}\left(\frac{v}{R}\right) f_0(u)\Theta(u - \Phi_w) \tag{17}$$

where $\Phi_w(x,t)$ is the wall potential relative to the axis, and $\Theta(x)$ is the step function. This expression assumes that only fast electrons with kinetic energies exceeding $\Phi_w$ escape to the wall. However, the calculation of $\Phi_w(x,t)$ may require two-dimensional analysis in configurational space $(x, r)$, and we plan to do that in future work.

In the present work, we utilized a simpler model previously used for the DC discharges [24]:

$$C_{wall} = -\langle C_{ion} \rangle_u \tag{18}$$

where $C_{ion}$ is defined by Eq. (14), and $\langle \quad \rangle_u$ includes energy integration in addition to space and AC-cycle time averaging $\langle \quad \rangle$ (see Eq. (24) below). This model assumes that all electrons are lost at an equal rate and ensures that the total number of particles in the computational domain remains constant [20].

For AC discharges, the averaging operator $\langle ... \rangle$ also included the time-averaging over an AC cycle. This model assumes that the electrons are lost at an average rate equal to the rate created over the striation length during the AC period. This model of average loss rate is appropriate for $\omega\langle\tau_a\rangle > 1$. Moreover, our tests showed that the choice of the electron loss model does not significantly affect the results for the reasons discussed above.

We introduce a computational grid in phase space for the numerical solution of the FP kinetic equation (15). The energy $u_{max}$ is selected about 2-3 $\varepsilon_1$. The typical number of cells in energy and space is 50 and 100. The boundary condition is specified as $f_0(x, u = u_{max}) = 0$ . The boundary condition at $u = 0$

$$\frac{\partial f_0}{\partial x} - E\frac{\partial f_0}{\partial u} \to 0 \tag{19}$$

ensures the absence of electron flux from the boundary at $u = 0$. The periodic boundary conditions in space for simulations of the positive column and the boundary conditions at electrodes for simulations of the entire discharge are discussed below.

Ions are described using a drift-diffusion model:

$$\frac{\partial n_i}{\partial t} + \frac{\partial}{\partial x}\left(\mu_i n_i E(x) - D_i \frac{\partial n_i}{\partial x}\right) = I - C_{wall} \tag{20}$$

where $\mu_i$ and $D_i$ are the ion mobility and diffusion coefficients, and $I$ is the ionization rate by electron impact. The electric field is calculated from the Poisson equation. We assume that the initial EEPF is Maxwellian, and the initial ion density equals the electron density.

The coupled set of the FP kinetic equation for electrons (15), the drift-diffusion for ions (20), and the Poisson equation for the electric field were solved using the finite element method in COMSOL with an implicit (BDF) time-stepping. This method is based on discretizing the simulation domain into elements inside which shape functions of different types and orders can be used. Fixed numbers of elements in both the coordinate (e.g., 100-200) and energy (e.g., 50-100) spaces were built out of quadrilateral mesh elements. We used the Lagrange shape-function type with linear first-order elements. Grid and element-order convergence studies were performed to verify computational accuracy. A direct (MUMPS) solver was employed at each time step for all quantities. We have not used the logarithmic transformation of the ion drift-diffusion and the FP kinetic equations used in our previous work [24]. It was found that the convergence of the coupled equations with different dimensionality, solved simultaneously, is substantially better without this transformation (significant for long transient simulations), and small negative values of $f_0$ and $n_i$ did not affect the results. Typical runs took several hours to days of wall-clock time on multi-core-enabled workstations. 5-10 core-based computations were found to be most numerically efficient.

## 3. Specifics of boundary conditions and radial loss model for AC positive column

The positive column is usually considered an autonomous system weakly influenced by near-electrode processes. Without striations, the positive column plasma is axially uniform with a local balance of the ionization and particle loss to the wall due to ambipolar diffusion or volume recombination. To analyze the stratification of the positive column, we apply periodic boundary conditions (BCs) for the EEPF and the particle fluxes at the left and right boundaries [24]. This BC ensures the continuity of the EEPF and the electron flux in phase space. We also apply periodic boundary conditions for ions by equalizing the ion density and flux at the boundaries.

Using periodic BCs for electrons and ions implies an integer number of waves fit the positive column. Also, such boundary conditions ensure that the total number of electrons and ions remains the same in the plasma (provided that matching electron and ion losses are used). As the total space charge remains zero (initially, a quasineutral plasma is assumed with $n_e = n_i$) along the striation length, the periodic BC for the electric field, $E_{\mathrm{left}} = E_{\mathrm{right}}$, is automatically satisfied. Therefore, we have prescribed zero potential at $x = 0$ and specified the potential drop $U\sin(2\pi\omega t) = U\sin(\tau)$ at $x = L$.

Similar to the DC striations studied previously [24], the tube radius $R$ was estimated as:

$$R = 2.4\sqrt{\frac{\langle D_a\rangle}{\langle \nu_i\rangle}} \quad , \tag{21}$$

where the averaged electron temperature $\langle T_e\rangle$ was used to evaluate the ambipolar diffusion coefficient, $D_a$, and spatial averaging over the striation length was calculated as:

$$\langle f\rangle(t) = \frac{1}{\Lambda}\int_0^{\Lambda} dx' f(x',t) \tag{22}$$

For AC discharge striations studied in the present paper, we have also added averaging over the AC period *T*:

$$\langle f\rangle = \frac{1}{\Lambda}\frac{1}{T}\int_{t-T}^{t}\int_0^{\Lambda} dt' dx' f(x',t') \tag{23}$$

This method allowed the tube radius to become spatially- and time-independent at periodic convergence. Due to current COMSOL limitations, "running" time averaging was used instead:

$$\langle f\rangle = \frac{1}{\Lambda}\frac{1}{t}\int_0^{t}\int_0^{\Lambda} dt' dx' f(x',t') \tag{24}$$

For computing the tube radius, oscillations of the electron temperature in the dynamic regime at $\omega\langle\tau_u\rangle < 1$, have been neglected.

# III. Standing striations in AC discharges

We obtained standing striations in the AC positive column with periodic BCs in Argon and Neon over a wide range of driving frequencies from 0.1 to 50 MHz. Our model resolved the AC period to compute the EEPF and the ion transport. We found that plasma density responds weakly to oscillating electric fields at driving frequencies at $\omega\langle\tau_a\rangle > 1$. On the other hand, at frequencies below ~25-50 MHz in Argon at p = 0.5 Torr, the EEPF and electron mean energy oscillated over the AC period because $\omega\langle\tau_u\rangle < 1$. The length of standing striations of a few tube radii depended weakly on driving frequency. Higher frequencies required longer simulation times.

## 1. Standing Striations in Argon

Figure 1 shows calculated spatial distributions of plasma properties in Argon at pressure $p$ = 0.4 Torr, plasma density $n_0 = 6 \cdot 10^8$ cm$^{-3}$, driving frequency of 1 MHz, and voltage $U$ = 55 V. The column length is $L$ = 6 cm, and the electric field has amplitude $E_0 = \frac{U}{L} = 9.17$ V/cm. Under these conditions, the calculated Debye length 0.07 cm is much smaller than the tube radius $R$ = 0.37 cm. Four standing striations with a length of $\Lambda = 1.5$ cm formed in our simulations. The time-averaged electron and ion densities deviate slightly near their max and min values. Substantial spatial oscillations of the time-averaged electron temperature are observed. For most of the AC period, the electric field reverses direction in space, forming collisional double layers typical of strongly nonlinear waves.

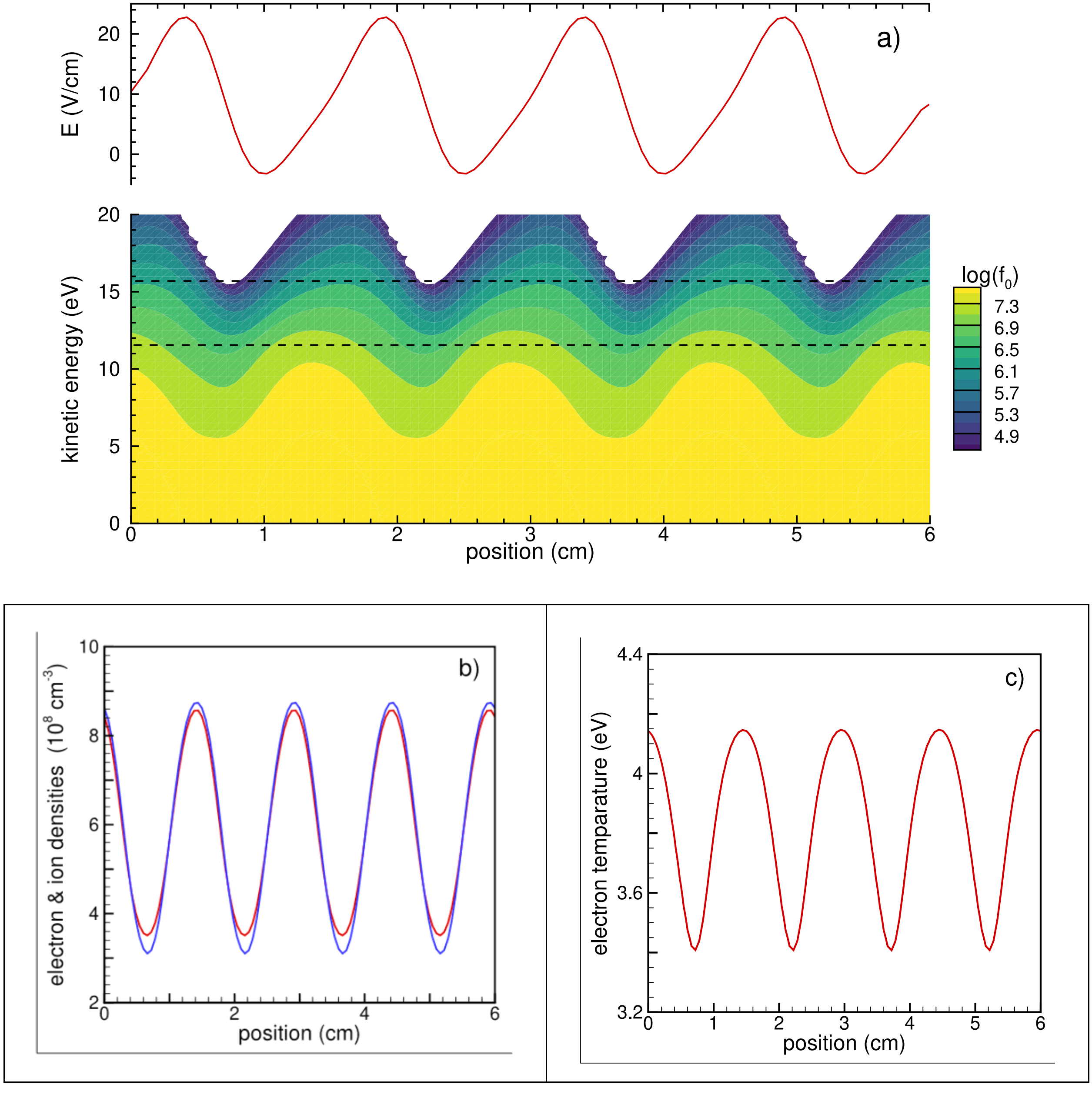

Figure 1: Instantaneous distribution of the electric field (a) and EEPF contours (dashed lines show the excitation and ionization thresholds). Time average electron and ion densities (b) and electron temperature (c), Argon, $p = 0.4$ Torr, 1 MHz, $E_0 = 9.17$ V/cm.

By changing the driving frequency and applied voltage, we could change the wavelength and modulation depth of the striations. The amplitude of the electric field, $E_0$, has the most substantial effect. Figure 2 shows results for 0.4 Torr, $n_0 = 10^9$ cm$^{-3}$, frequency of 5 MHz, and $U = 80$ V. These conditions correspond to $E_0 = 13.3$ V/cm, the tube radius of $R = 3.2$ mm, and the Debye length of about 0.5 mm. The striation length is $\Lambda = 1.6$ cm. As in the previous case of 1 MHz, the discharge operates in a dynamic regime. The ion and electron densities remain unchanged during the AC period, but the electron temperature and the ionization rate oscillate substantially around their average values. We refer to these conditions as a dynamic regime of discharge operation. The amplitude of modulations for these quantities decreases gradually with increasing driving frequency.

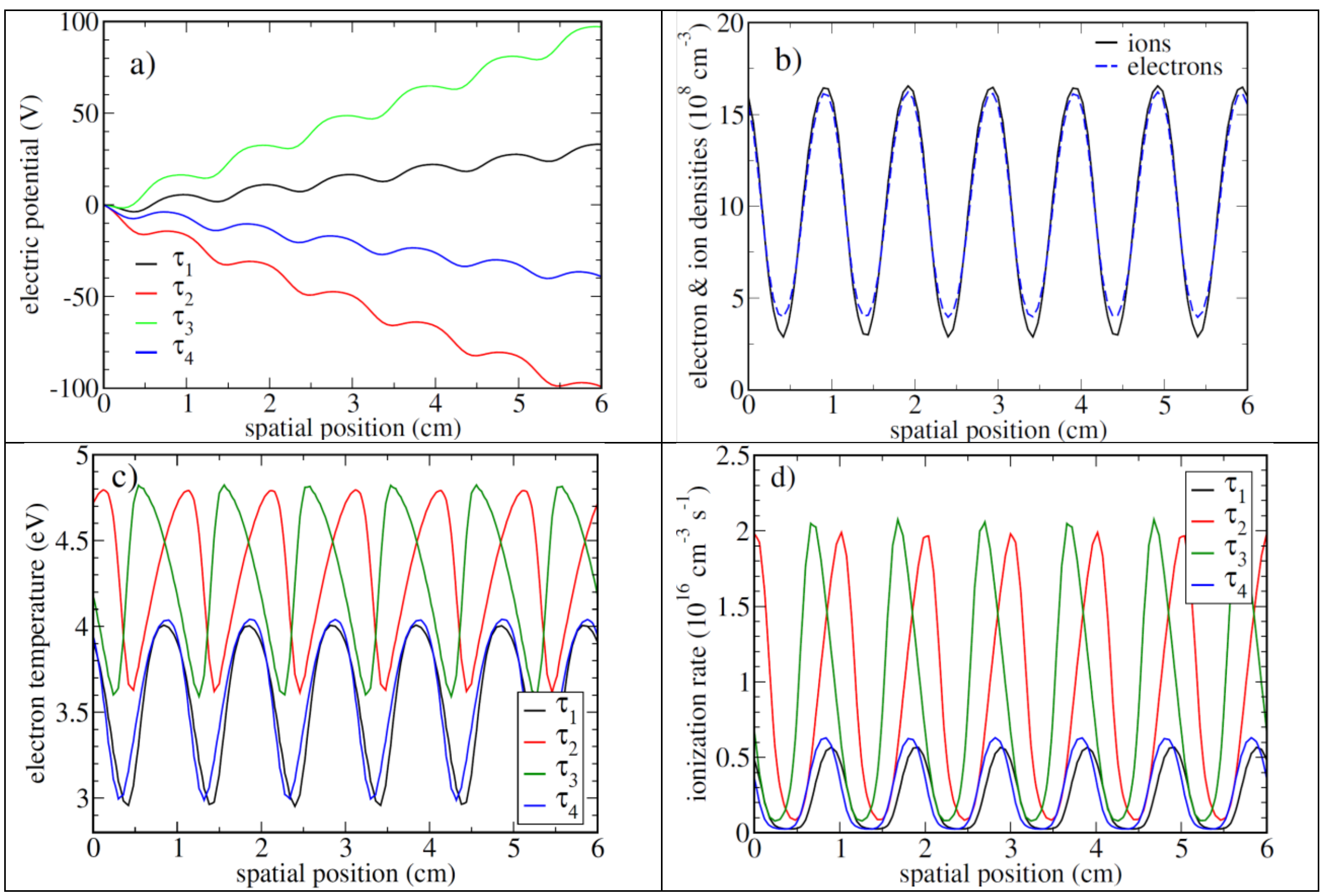

Figure 2: Spatial distributions of the electric potential (a), electron and ion densities (b), electron temperature (c), and the ionization rate (d) at four phases $\tau_1 = 3\pi/4$, $\tau_2 = 5\pi/4$, $\tau_3 = \pi/2$, and $\tau_4 = 3\pi/2$ during the AC period. Argon, 0.4 Torr, 5 MHz, and $E_0 = 13.3$ V/cm.

Figure 3 shows an example of temporal variations of plasma properties in Argon at 0.4 Torr, at frequency 25 MHz, $U = 100$ V, $n_0 = 10^9$ cm$^{-3}$, $L = 3$ cm. This case corresponds to the large electric field of amplitude $E_0 = 33.3$ V/cm, small tube radius of $R = 1$ mm, and Debye length of about 0.5 mm, comparable to the tube radius. The striation length is $\Lambda = 6$ mm. Substantial differences in electron and ion densities occur at the positions of their maximum and minimum values (b). The time modulations of electron and ion densities are negligible. However, substantial modulations of electron temperature (c) and the rates of electron-induced inelastic processes (at a double AC frequency like those shown in Figure 2) are observed. These modulations are associated with oscillations of the EEPF at low and high energies (d). The "body" of the EEPF remains practically unchanged during the AC period. The EEPF "tail" is enhanced due to heating fast electrons, whereas its "head" is depleted by low energy electrons, so the total electron density remains unchanged. The depletion of low-energy electrons corresponds to increasing mean energy (temperature). Recall that low-energy electrons control the dynamics of heat transport processes, whereas fast electrons control electron-induced excitation and ionization processes.

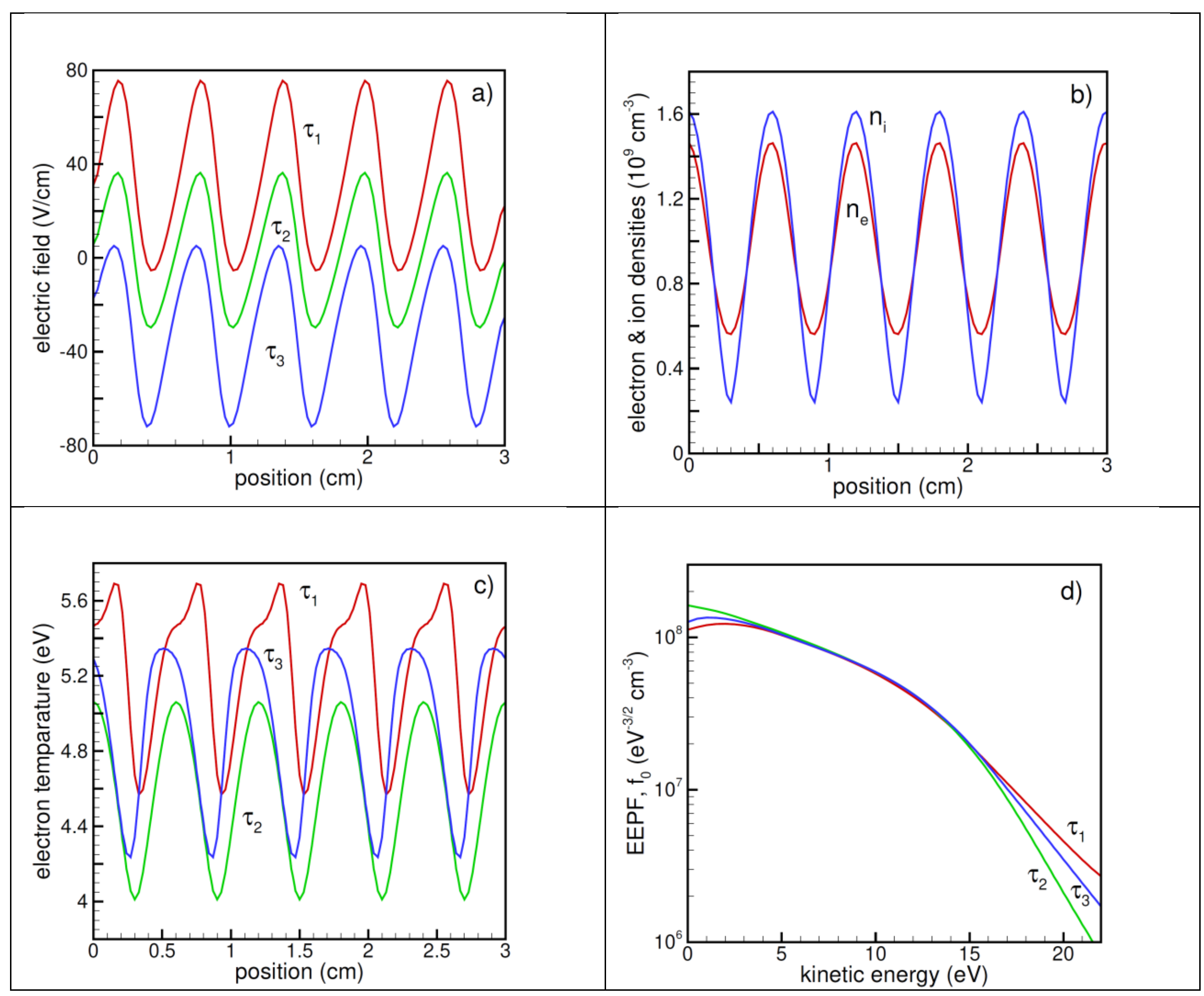

Figure 3: Instantaneous spatial distributions of the electric field (a), electron and ion densities (b), electron temperature (c), and EEPF (d) at three phases $\tau_1 = \pi/2$, $\tau_2 = \pi$, and $\tau_3 = 3\pi/2$ during AC period. Argon, 0.4 Torr, 25 MHz, $E_0 = 33.3$ V/cm.

To clarify the effects of gas pressure and driving frequency, we show below our results for $p = 0.1$ Torr and driving frequencies 0.1 and 50 MHz. Figure 4 shows the results for 0.1 MHz, $n_0 = 10^{10}$ cm$^{-3}$, and $U = 55$ V, for the column length of $L = 24$ cm. The electric field amplitude is $E_0 = 1.86$ V/cm, the calculated tube radius is $R = 5$ cm, and the striation length is $\Lambda = 8$ cm. For most of the AC period, the electric field does not change sign in space (a), indicating low-amplitude waves. Noticeable spatial movements of the electron and ion densities over their average values are observed during the AC period (b). Strong oscillations of electron temperature in space and time are observed (c) at this low driving frequency and low-pressure case compared to those described above. This case is at the boundary of the applicability of our model, which assumes $\omega\langle\tau_a\rangle > 1$.

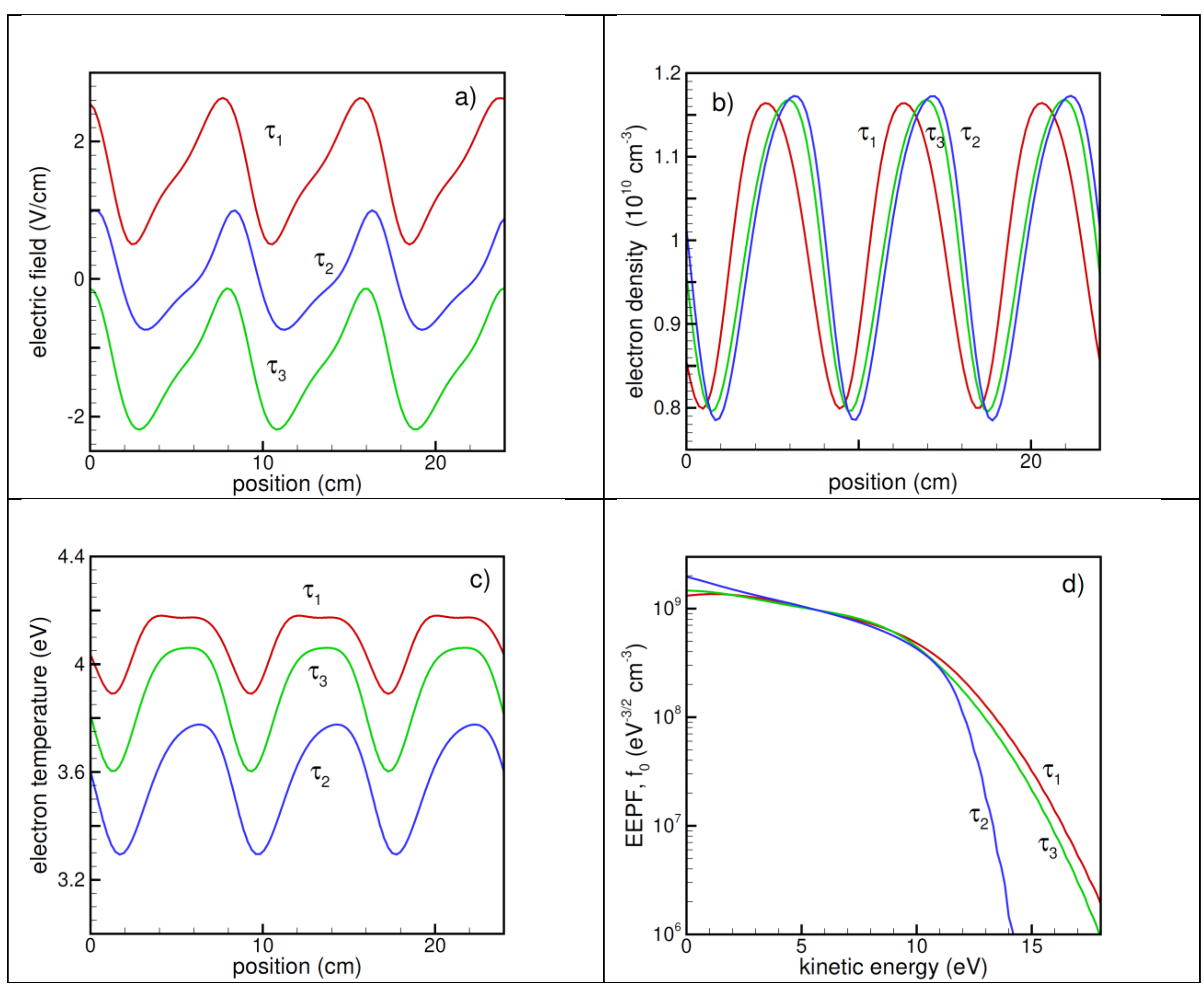

Figure 4: Instantaneous spatial distributions of the electric field (a), electron density (b), electron temperature (c), and EEPF at $x$ = 4 cm at three phases $\tau_1 = \pi/2$, $\tau_2 = \pi$, and $\tau_3 = 3\pi/2$ during the AC period. Argon, 0.1 Torr, 0.1 MHz, $E_0 = 1.86$ V/cm.

Figure 5 shows results for the driving frequency of 50 MHz, $n_0 = 10^{10}$ cm$^{-3}$, and voltage $U$ = 45 V, for the column length of $L$ = 6 cm. Three standing striations formed with a length of $\Lambda = 2$ cm. The amplitude of the electric field is $E_0 = 7.5$ V/cm. However, the ambipolar field created in stratified plasma exceeds the applied AC field (a). Under these conditions, plasma is strictly quasineutral, as the calculated Debye length of 0.3 mm is much smaller than the tube radius $R$ = 0.52 cm. The densities of electrons and ions are not distinguishable (b) and do not change over the AC period. However, oscillations of the electron temperature around their average value are still present (c). These oscillations are associated with substantial variations of the low-energy part of the EEPF (d). Small oscillations of the EEPF tail result in the temporal variation of excitation and ionization rates (like those illustrated in Figure 2c). Electron heat transport occurs at the rate of free electron diffusion, i.e., much faster than ambipolar diffusion, which controls the plasma density profile. This effect explains why the temporal variations of electron temperature in spatially non-uniform plasma are observed even for such high driving frequencies.

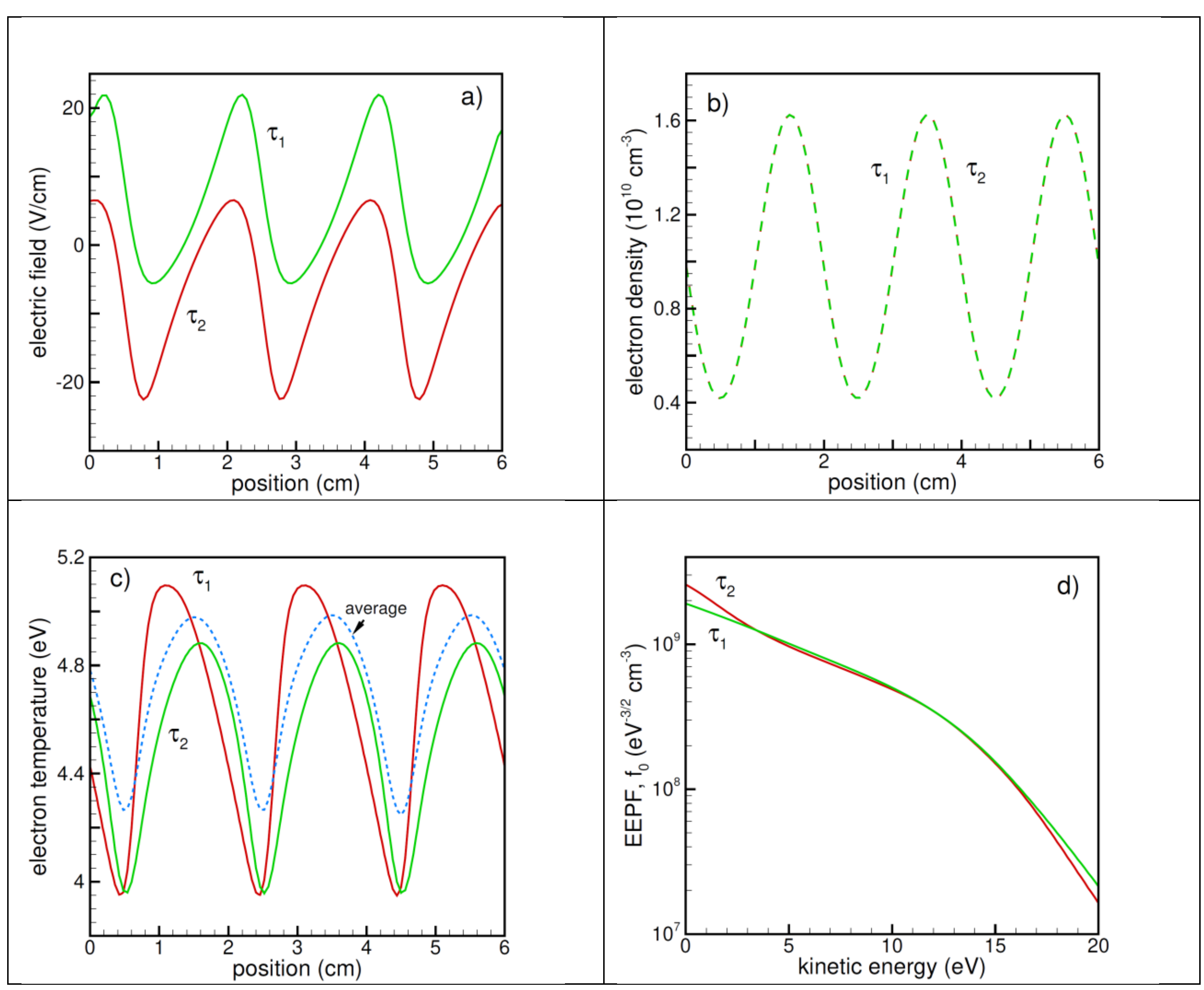

Figure 5: Instantaneous spatial distributions of the electric field (a), electron temperature, electron density (b), electron temperature (c), and the EEPF at a point $x = 2$ cm for two phases $\tau_1 = \pi/2$, and $\tau_2 = 3\pi/2$ during the AC period. Argon, 0.1 Torr, 50 MHz, $E_0 = 7.5$ V/cm.

Figure 6 compares temporal variations of the electric field, electron temperature, and electron density in the middle of the column for 0.1 and 50 MHz. Substantial oscillations of the plasma density are observed at 0.1 MHz, whereas the density oscillations at 50 MHz are negligible. The temperature oscillations at 0.1 MHz occur at the double driving frequency, whereas at 50 MHz, they occur at the driving frequency. The difference in the absolute values of the electric field at the observation point is due to the different positions of this point to density maximum in the striation.

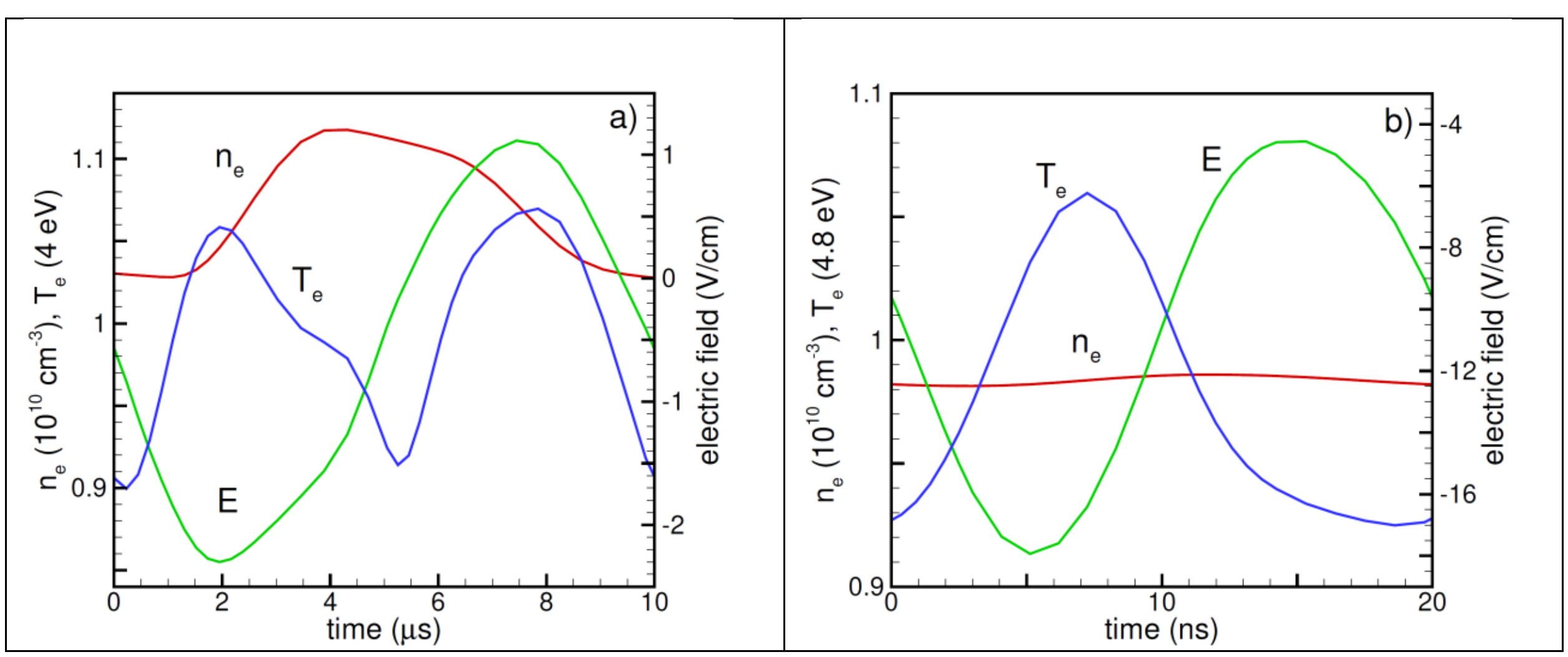

Figure 6: Temporal variations of the electric field (green), electron temperature (blue), and electron densities (red) at L/2 for 0.1 MHz (a) and 50 MHz (b) in Argon at $p$ = 0.1 Torr.

Figure 7 summarizes the dependence of striation length $\Lambda$ on the electric field amplitude $E$ and tube radius $R$ observed in our simulations. For a wide range of electric fields, the $\Lambda(E)$ dependence fits well with the formula $\Lambda = a/E$, where $a$ = 15.7 eV (Figure 7a). According to Novak's law, the expected value of $a$ should be between the excitation and ionization thresholds, which are $\varepsilon_1$ = 11.55 and $\varepsilon_i$ = 15.8 eV for Argon. Therefore, the observed dependence of the striation length on the electric field in Argon AC discharges obeys Novak's law. It is seen in Figure 7 (b) that the striation length is about a few tube radii, which generally agrees with experiments.

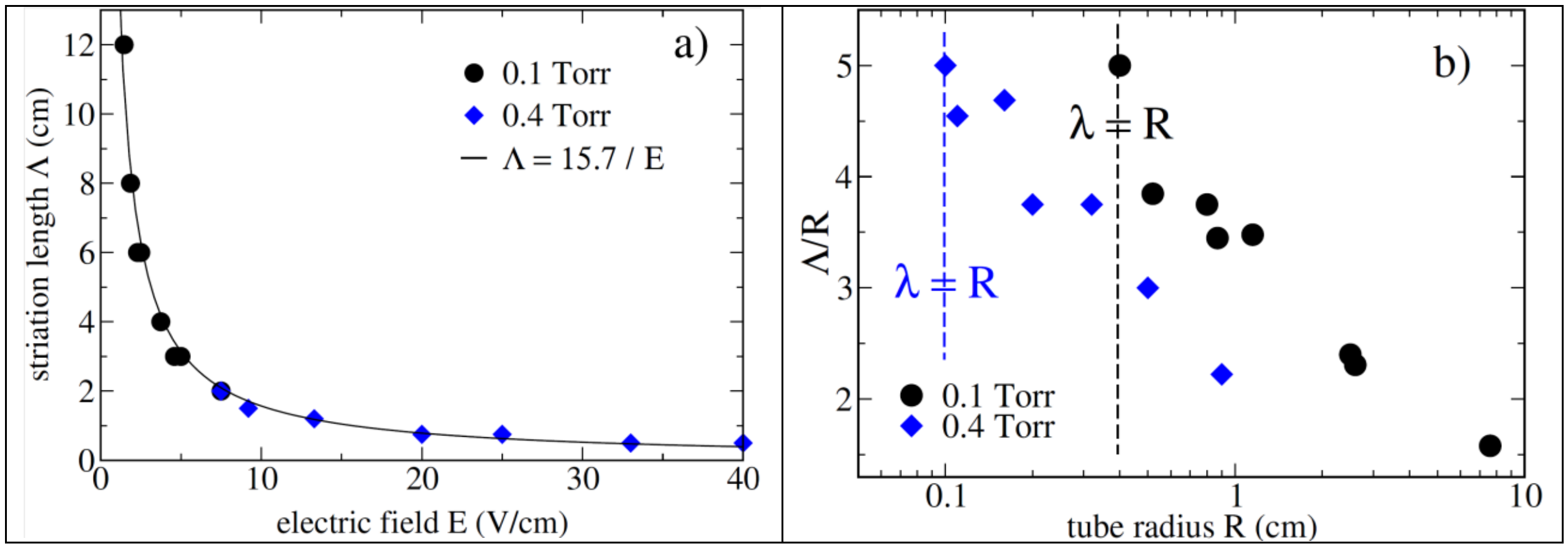

Figure 7: Striation length vs. the electric field (a) and tube radius (b) in Argon.

## 2. Standing Striations in Neon

Our simulations for Neon also produced standing striations at $p$ = 1 Torr, electric fields 3 - 15 V/cm, and driving frequencies of 0.1, 1, and 10 MHz. Figure 8 shows results for 1 MHz, $n_0$ = $10^{15}$ m$^{-3}$, and $U$ = 30 V, for the column length $L$ = 10 cm. For this low value of the electric field $E_0 = 3$ V/cm, our simulations produced low-amplitude striations with a length of $\Lambda = 5$ cm, for the tube radius $R$ = 3.3 cm (see Figure 8a). The ion and electron densities are practically constant over the AC period (b). However, substantial temporal and spatial oscillations of electron temperature are observed (c). The EEPF body remains practically unchanged during the AC period (d), whereas its head and especially tail demonstrate substantial temporal dependencies.

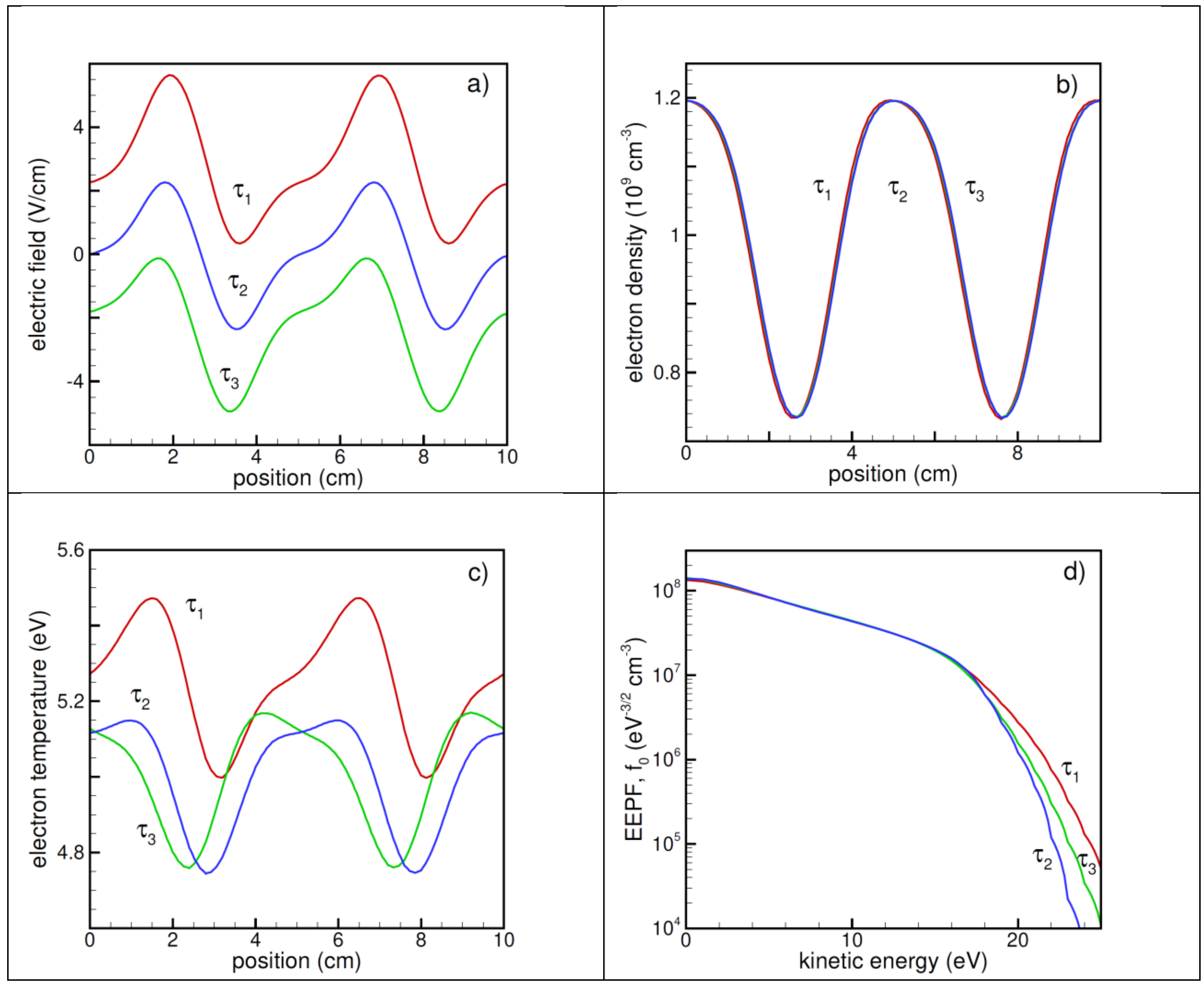

Figure 8: Time and space resolved electric field (a), electron density (b), electron temperature (c), and EEPF (d) at three phases $\tau_1 = \pi/2$, $\tau_2 = \pi$, and $\tau_3 = 3\pi/2$ during the AC period. Neon, 1 MHz, $E_0 = 3$ V/cm.

Figure 9 shows results for 0.1 MHz, $n_0 = 10^{15}$ m$^{-3}$, and $U$ = 50V. The electric field amplitude is $E_0$ = 5 V/cm, the calculated tube radius is $R$ = 1.6 cm, and Debye length is 1 mm. The positions of maximum and minimal values of the electric field are shifted in space (a). As a result, two maximums of the election temperature per striation length tend to form (see the $\tau_2$ curve in Figure 9 (c)). The EEPF appears to have a three-temperature shape (d) and oscillates noticeably over the AC period.

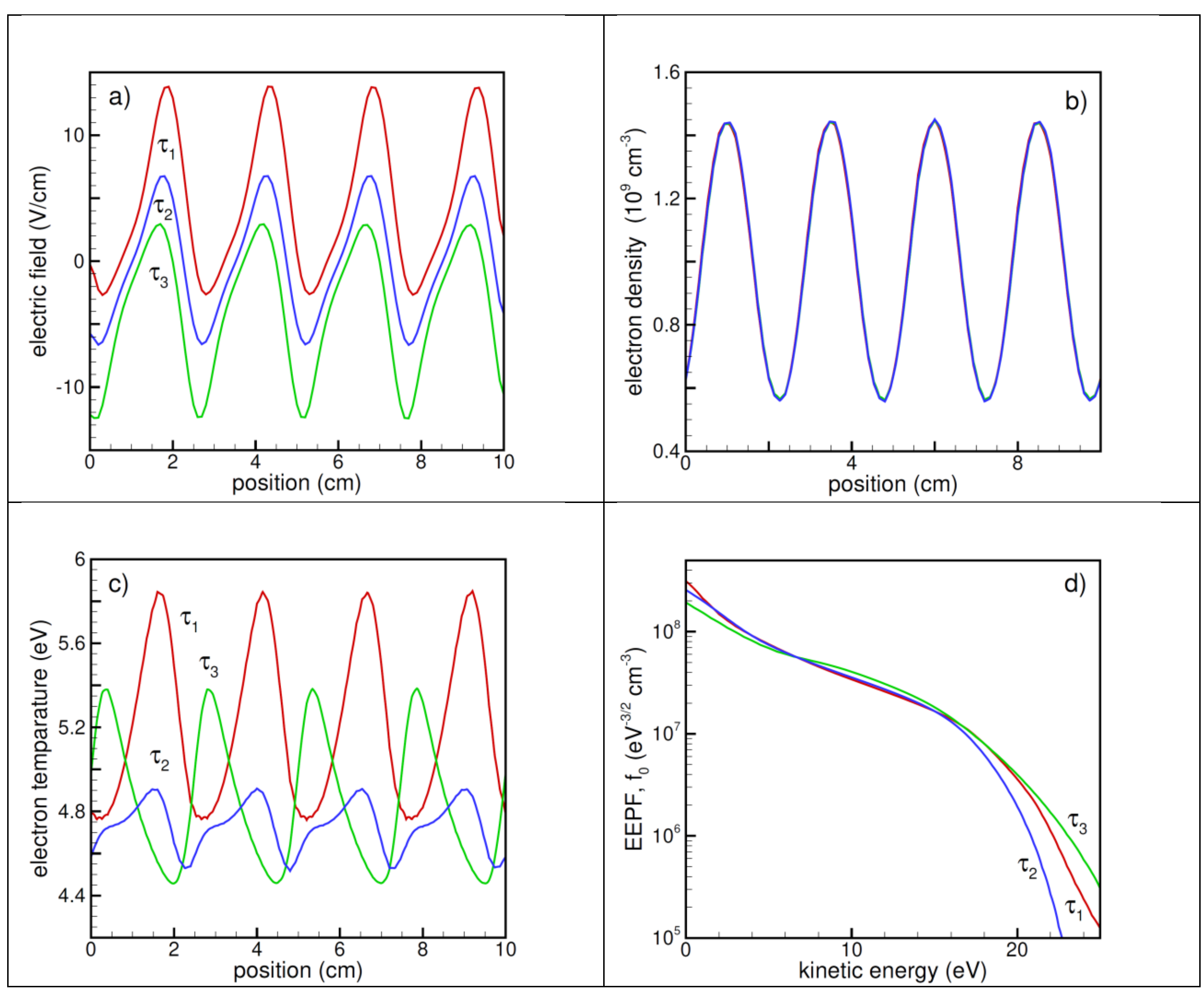

Figure 9: Time and space resolved electric field (a), electron density (b), electron temperature (c), and EEPF at $x = 3$ cm (d) at three phases $\tau_1 = \pi/2$, $\tau_2 = \pi$, and $\tau_3 = 3\pi/2$ during AC period. Neon, 0.1 MHz, $E_0 = 5$ V/cm.

Figure 10 compares temporal variations of the electric field, electron temperature, and electron densities in the middle of the column for 0.1 and 1 MHz. At 0.1 MHz, small oscillations of plasma density are observed, whereas the density oscillations at 1 MHz are negligible. The temperature oscillations at 0.1 MHz are substantial and occur at double the driving frequency, whereas, at 1 MHz, they are already small. The absolute values of the electric field at the observation point depend on the position of this point to the maximum plasma density in the striation.

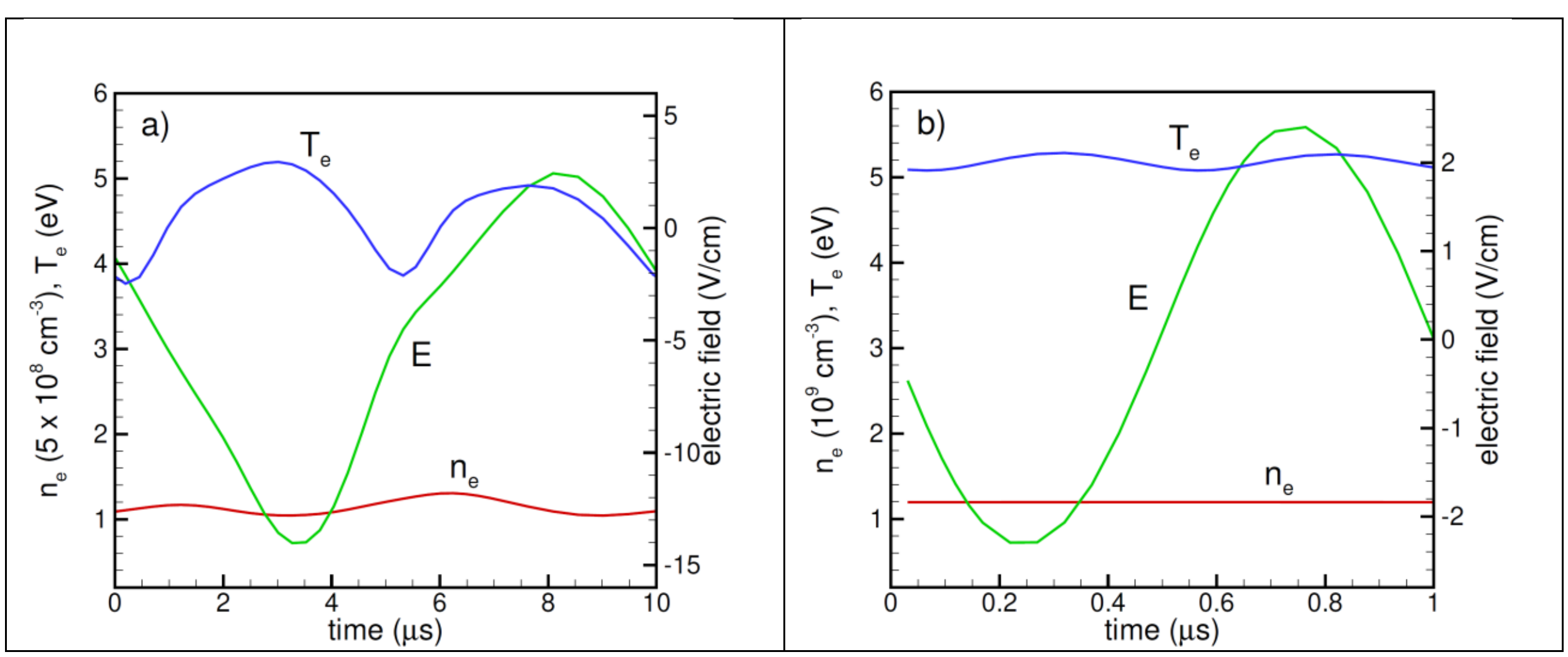

Figure 10: Temporal variations of the electric field (green), electron temperature (blue), and electron densities (red) at $x = L/2$ for 0.1 MHz (a) and 1 MHz (b) in Neon at $p$ = 1 Torr.

Figure 11 shows the dependence of the striation length $\Lambda$ on the electric field $E$ and the tube radius $R$ obtained in our simulations for Neon. The calculated $\Lambda(E)$ dependence can be approximated by the formula, $\Lambda = a/E$, with the expected value of $a$ between the excitation and ionization thresholds, $\varepsilon_1$ = 16.62 and $\varepsilon_i$ = 21.6 eV for Neon. Therefore, in our calculations, $\lambda_\varepsilon < \Lambda < \lambda_{\varepsilon_i}$, in accordance with Novak's law. The $\Lambda(E)$ dependence in Figure 11a is better approximated when the $a$'s value increases with increasing $E$, which appears reasonable considering a simplified two-level Neon atom with only direct ionization channel. The striation length $\Lambda$ is about three tube radii $R$, shown in Figure 11b, which generally agrees with experiments.

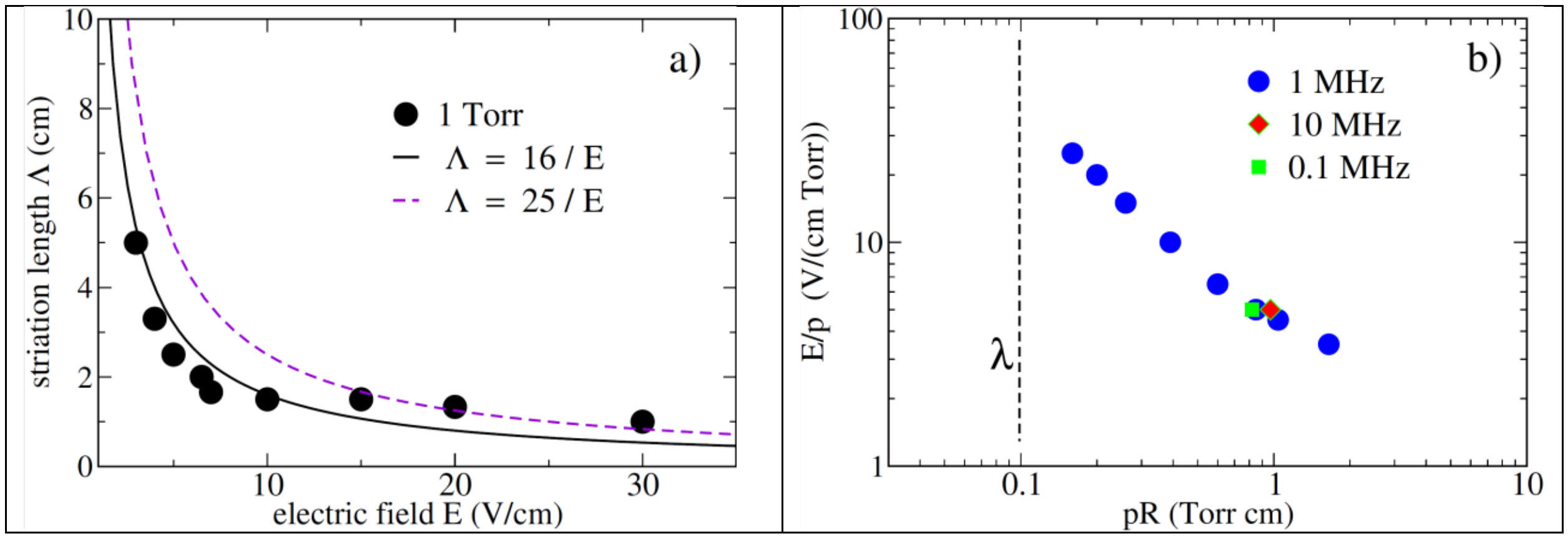

Figure 11: Calculated striation length vs. the electric field (a) and the electric field vs tube radius (b) in Neon at $p$ = 1 Torr.

The dashed line in Figure 11b shows the electron mean free path $\lambda$ at 1 Torr. For Neon, the transport collision cross section can be assumed energy independent, $\sigma_{tr} = 4 \cdot 10^{-16}$ cm. Therefore, $\lambda = 1/(N\sigma_{tr}) = 0.1$ cm and $\lambda_u = \frac{\lambda}{\sqrt{\delta}} = 14$ cm at 1 Torr. Since $\lambda < R$ in our simulations, we are within the applicability of two-term SHE. Furthermore, the range of $E/p$ in our simulations corresponds to the inelastic energy balance for electrons ($\lambda_u > \lambda_\varepsilon$). Our studies were limited to low plasma densities when non-linear effects associated with stepwise ionization and Coulomb collisions among electrons are negligible.

## 3. Mechanism of plasma stratification in AC positive column

Our simulations described above demonstrate that the electric field in AC discharge can be separated into ambipolar and conduction components, $\boldsymbol{E}(\boldsymbol{r},t) = \boldsymbol{E}_a(\boldsymbol{r},t) + \boldsymbol{E}_c(\boldsymbol{r},t)$, which evolve at different time scales. The ambipolar field is controlled by slow ion motion and evolves at the slow (ambipolar diffusion) time scale. This field can be written as $\boldsymbol{E}_a = -\nabla\varphi$ , where $\varphi(\boldsymbol{r},t)$ is the electrostatic potential [30]. The conduction component evolves at the time scale associated with driving frequency $\omega$. The conduction electric field, $\boldsymbol{E}_c(\boldsymbol{r},t)$ describes electric current in plasma and electron heating.

In this section, we calculate the electric field in plasma from the current conservation [6]:

$$E(x,t) = \frac{j(t)}{eb_e n_e(x)} - \frac{T_e(x,t)}{en_e}\frac{\partial n_e}{\partial x} = E_c(x,t) + E_a(x,t) \tag{25}$$

Here, $E_c(x,t)$ is a resistive component responsible for the conduction current and $E_a(x,t)$ is the ambipolar field responsible for the plasma quasineutrality. The field separation model has been previously used for simulations of the AC positive column where the axial (resistive) and radial (ambipolar) electric fields are orthogonal [28]. The field separation model is commonly used for ICP at high driving frequencies but has been rarely applied for CCP because it may not be applicable in the sheaths (see Section IV below).

We have seen in the previous section that in the range of driving frequencies, $\omega\langle\tau_a\rangle > 1$, the ambipolar field oscillates slightly over time because the plasma density remains quasi-steady. The time oscillations of the ambipolar electric field at frequencies $1/\langle\tau_a\rangle < \omega < \nu_u$ occur mainly because of electron temperature variations rather than density variations (this regime was labeled the dynamic regime).

$$E_a(x,t) = -\frac{T_e(x,t)}{en_e}\frac{\partial n}{\partial x} \tag{26}$$

The expression (26) for the resistive component neglects the temporal dispersion of plasma conductivity, which is appropriate for $\omega \ll \nu$ . The latter can be easily included [18] and has been commonly used for ICP simulations [31].

In our simulation described in this section, the ambipolar field, $E_a(x)$, was computed by solving the Poisson equation with fixed zero potential at both ends of the column. The $E_c(x,t)$ was computed from Eq. (26) by specifying the discharge current $j(t)$ and neglecting the time variations of $n_e$, which is valid at $\omega\langle\tau_a\rangle > 1$:

$$E_c(x,t) = \frac{j(t)}{eb_e n_e(x)}$$

This field separation model makes the analysis of electron kinetics in AC plasma more transparent. Using total energy $\varepsilon = u - \varphi(x,t)$ as an independent variable, the kinetic equation (9) can be rewritten as:

$$\frac{\partial f_0}{\partial t} - \frac{\partial}{\partial x}\left(D_r \frac{\partial f_0}{\partial x}\right)_\varepsilon - \frac{1}{\sqrt{\varepsilon}}\frac{\partial}{\partial \varepsilon}\left(\sqrt{\varepsilon} D_u(x,t)\frac{\partial f_0}{\partial \varepsilon}\right) = S_0 \tag{27}$$

The left part of Eq. (28) describes electron diffusion over surfaces of constant total energy, $\varepsilon = u - \varphi(x,t)$ in phase space $(x,u)$, and diffusion over $\varepsilon$ due to heating with a diffusion coefficient $D_u(u,x,t)$, which is proportional to $E_c^2(x,t)$.

We assume mono-harmonic conduction current, $j(t) = j_0\, cos(\omega\, t)$, and the electric field $E_c(x,t) = E_0\,(x) cos(\omega\, t)$. By taking the integral (7), we obtain [35]:

$$D_u(u,x,t) = \frac{1}{2} D_0(u)\left\{(1+\cos(2\omega t))\left(\frac{\nu^2}{\omega^2+\nu^2}\right) + \sin(2\omega t)\left(\frac{\nu\omega}{\omega^2+\nu^2}\right)\right\} \tag{28}$$

where $D_0(u,x) = {E_0}^2(x) D_r(u)$. At $\omega \ll \nu$, this formula corresponds to a quasi-static case:

$$D_u(x,t) = D_0(u,x)\cos^2(2\omega t) \tag{29}$$

At $\omega \gg \nu_u(u)$, the EEPF $f_0(x,u,t)$ is controlled by the time average field:

$$D_u(u,x) = D_0(u,x)\frac{1}{2}\frac{\nu^2}{\nu^2+\omega^2} \tag{30}$$

The time scales for temporal EEPF relaxation in the elastic and inelastic energy ranges, at $u < \varepsilon_1$ and $u > \varepsilon_1$ differ considerably because the frequency $\nu_u(u)$ depends strongly on electron kinetic energy in noble gases.

Figure 12 shows the results of simulations for a frequency of 50 MHz at Argon pressure of 0.4 Torr, and $j_0$ = 20 mA/cm$^2$. Under these conditions, Eq. (31) is applicable. The calculated radius $R$ = 3.7 mm, the electric field is $E_0$ = 15 V/cm, the Debye length is 0.2 mm, and the striation length is $\Lambda = 1$ cm. The electron density $n_e(x)$ are maximal at the minimal electric potential $-\varphi(x)$ , see Figure 12 (a), as typical for plasma controlled by ambipolar diffusion.

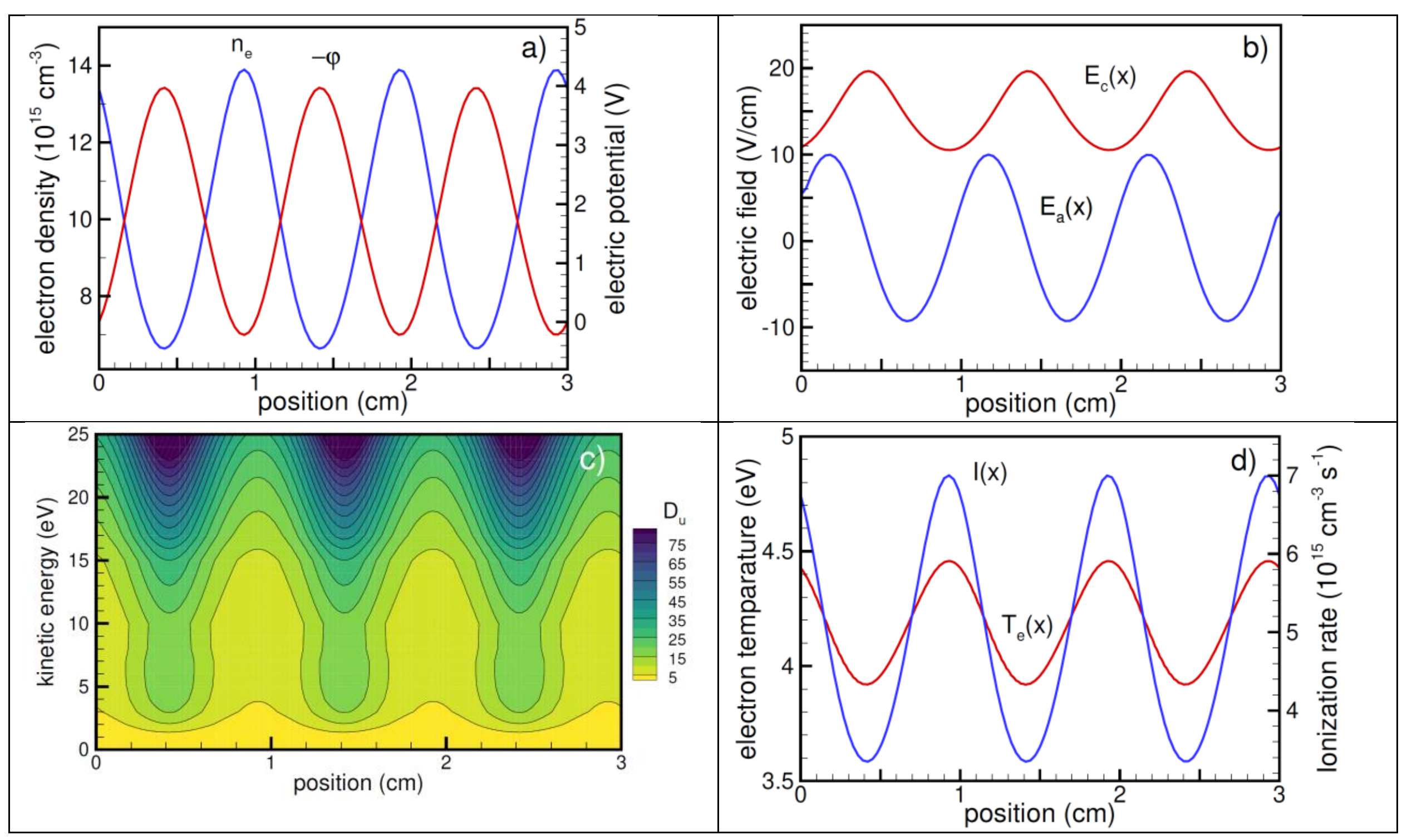

Figure 12: Spatial distributions of the electron density and ambipolar potential (a), resistive and ambipolar electric fields (b), diffusion coefficient (c), electron temperature, and ionization rate (d) in Argon at 50 MHz, 0.4 Torr, $j_0 = 20$ mA/cm$^2$.

The resistive and ambipolar fields, $E_0(x)$ and $E_a(x)$, have comparable amplitudes (b). The maximal values of the diffusion coefficient $D_u(u,x)$ are at the positions of minimal plasma density, where $E_0(x)$ has maximal values (c). The energy dependence of the diffusion coefficient $D_u(u,x)$ is determined by the behavior of elastic collision frequency. As the diffusion coefficient increases with energy for Argon, the heating of low-energy electrons is suppressed. The electron temperature and the ionization rate obtained in our simulations have maximum values near the points of maximal electron density (d), which supports plasma stratification.

The most favorable conditions for stratification occur when the ambipolar and resistive components have comparable amplitudes. The second and third terms in Eq. (27) are of the same order at $\Lambda \approx \varepsilon_1/(e\langle E_{eff}\rangle)$, where $\langle E_{eff}\rangle = U/L/\sqrt{2}$ . Therefore, the intrinsic spatial scale $\lambda_\varepsilon = \varepsilon_1/(e\langle E\rangle)$ determines the wavelength of striations in both DC and SAC discharges, where $\langle E\rangle = U/L$ for the DC case and $\langle E\rangle = U/L/\sqrt{2}$ for the AC case.

## IV. Effects of electrodes on plasma stratification

We have also performed simulations of CCP with resolved RF sheaths. The electron emission from electrodes was neglected, which corresponds to the $\alpha$ mode of CCP operation. Figure 13 shows an example of simulations in Neon at $p = 0.5$ Torr, an interelectrode gap of 20 cm, a driving voltage of 300 V, and a frequency of 3 MHz. In this case, the tube radius of $R = 1$ cm was prescribed, and the code calculated the electric field in the plasma column. Nonlinear standing striations with a wavelength of about 4 cm formed inside the gap (a). Plasma density has maximum values at the minimum time-average electric potential (b). The depth of the potential well for electrons is lower by about 30 V than the amplitude of the applied voltage. The electron and ion densities respond to the time average electric field in the plasma; they do not change substantially over the RF period. Ions also react to the time-average electric field inside the sheath, whereas electrons respond to the instantaneous field (see electron dynamics in the sheaths in Figure (a). The electron temperature oscillates substantially in time (and space) in plasma (d), corresponding to the dynamic discharge operation regime.

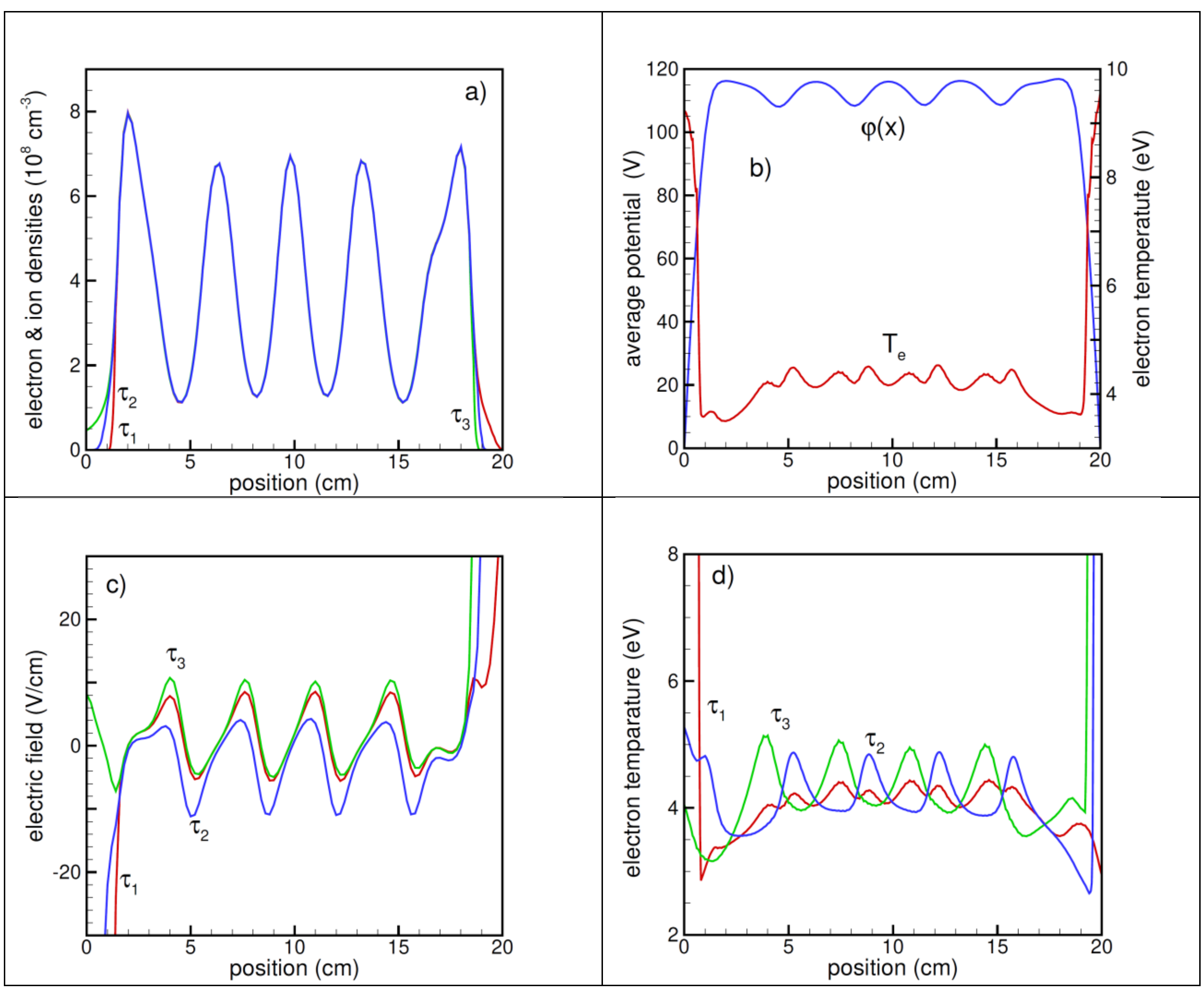

Figure 13: Instantaneous spatial distributions of electron density (a), period-average distributions of the electric potential and electron temperature (b), electric field (c), and electron temperature (d) at three phases $\tau_1 = \pi/2$, $\tau_2 = \pi$, and $\tau_3 = 3\pi/2$ during the AC period.

The main difference between the striations with periodic BCs is a non-sinusoidal time dependence of the electric field, $E(t)$ in the plasma due to the nonlinear processes in the RF sheath (Figure 14). The analysis of particle kinetics in the sheath and near-electrode effects is beyond the scope of the present paper. In our simulations, we saw that any asymmetry of discharge induced a DC electric field and slow motion of striations, as was previously observed in experiments [6].

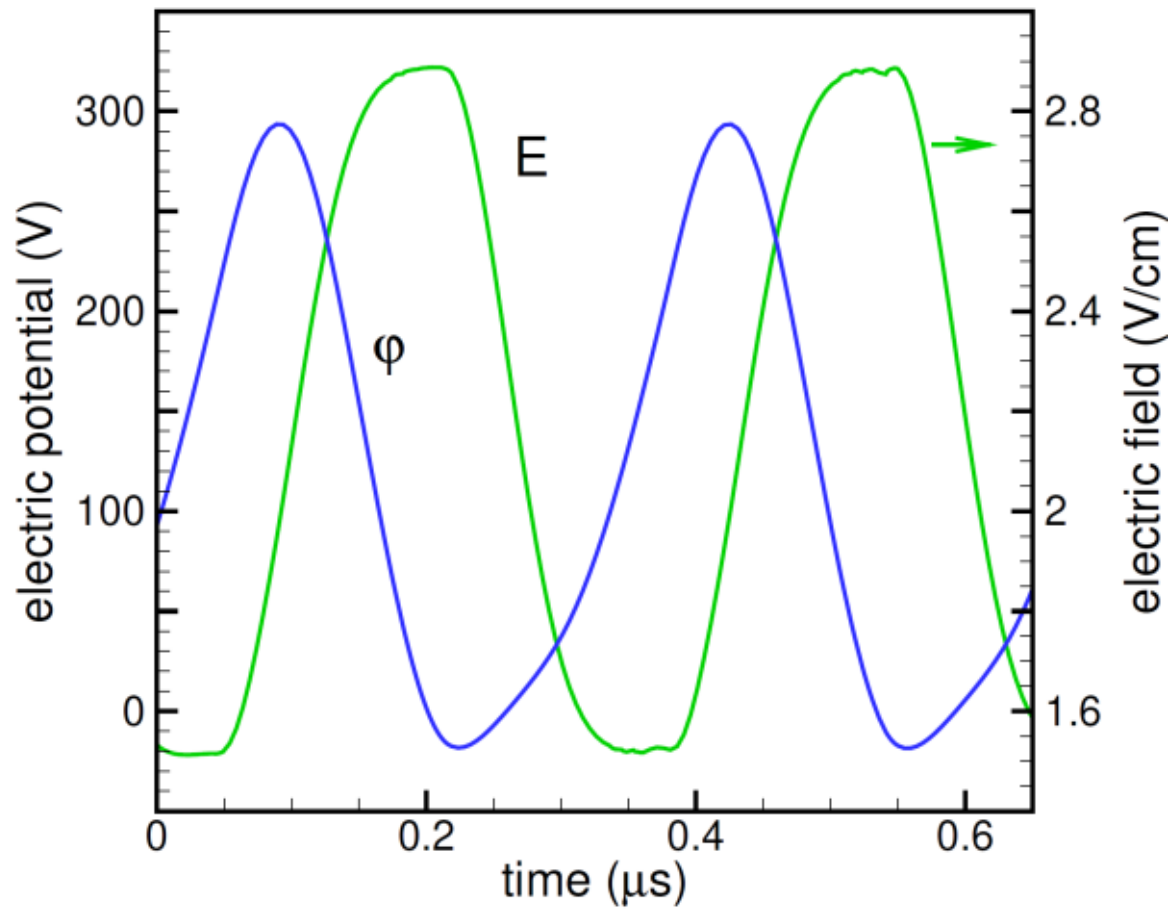

Figure 14: Time variations of the electric field and potential in the middle of the gap.

We emphasize that in this section, the tube radius was prescribed rather than calculated by the code as in Section III for periodic boundary conditions. Indeed, the model with electrodes is closer to the experimental settings. However, Section III's results are essential for studying striations' resonance properties, as described in detail in our previous paper [24] devoted to moving striations in DC discharges. This is discussed in the following section.

## V. Discussion

We conducted self-consistent simulations of standing striations in AC discharges of noble gases with a hybrid model accounting for nonlocal electron kinetics. From the general theory [26], it is known that the spatial non-locality of the EEPF is associated with two intrinsic spatial lengths: $\lambda_\varepsilon = \varepsilon_1/(eE)$ over which the electrons gain kinetic energy equal to the excitation threshold $\varepsilon_1$, and $\lambda_u = \lambda(\nu/\nu_u)^{1/2}$ the energy relaxation length for electrons in collisions. The minimum of these lengths defines striation length in stratified plasma. At $\lambda_\varepsilon < \lambda_u$, which corresponds to the inelastic energy balance of electrons, several types of striations with lengths close to integer fractions of $\lambda_\varepsilon$ appear in DC discharges at low currents. In the opposite case, $\lambda_\varepsilon > \lambda_u$ , which corresponds to the elastic energy balance of electrons, one type of wave with a length of $\Lambda \approx \lambda_u$ has been observed. We have shown that our hybrid model, including the FP kinetic solver for electrons, describes the experimentally observed standing striations in AC discharges at low currents, with the wavelength controlled by $\lambda_\varepsilon$. Our simulations described above found that electron heating occurs mainly at minimums of electron density in stratified plasma columns. This behavior makes the stratified plasma column resemble the previously studied ICP and CCP discharges, where electron heating occurs predominantly near plasma boundaries where electron density is low [25].

Previous works on AC discharge stratification considered the thermocurrent instability predicted by Timofeev [18] for wavelengths longer than the energy relaxation length of electrons in elastic collisions at $\Lambda > \lambda_u$. Timofeev used a fluid model for electrons without an energy balance equation, which is justified when the local value of *E/N* controls a non-Maxwellian EEPF. He considered both the DC and the high-frequency case, $\omega \gg \delta\nu$, and thus first predicted the thermocurrent instability as a possible mechanism of stratification for high-frequency discharges. The thermocurrent instability appeared when the thermodiffusion electron flux exceeded the diffusion flux. However, Timofeev considered the conditions of the elastic energy balance of electrons and neglected inelastic collisions and ionization processes. Shveigert [36] conducted numerical instability analysis for Helium DC discharges using the kinetic equation for electrons with account for inelastic collision and direct ionization by electron impact. He confirmed that thermocurrent instability in Helium occurs at *E/N* = 4 - 7 Td. Shveigert observed that switching off the ionization changed the instability increment at *E/N* < 7 Td unsubstantially. Dyatko et al. [37] conducted a kinetic study of the linear stage of thermocurrent instability in DC electric fields and showed that the increment of this instability has a maximum at $\Lambda \approx \lambda_u$. They emphasized that the fluid description of electrons is invalid under these conditions. These previous works could not identify specific conditions and clarify the importance of ionization and heat transport processes in forming standing striations in AC discharges.

Fluid models with added energy balance for electrons have been suggested to capture nonlocal kinetic effects [38], [39]. Such models are well justified for noble gases at large plasma densities (currents), i.e. at $\Lambda > \lambda(\nu/\nu_{ee})^{1/2}$, where $\nu_{ee}$ is the frequency of Coulomb collisions among electrons. They have been successfully used in [6,12] to explain striations in DC and RF discharges in noble gases. One can also justify such fluid models for molecular gases where excitations of the vibrational state of molecules shorten the energy relaxation length of electrons [40]. However, extended fluid models have been recently used to explain plasma stratification in DC [20] and RF [16,17] discharges of noble gases at low plasma densities. Unsurprisingly, such models could describe some features of stratification in noble gases by capturing some nonlocal effects. However, they fail to explain Novak's law and the kinetic resonances responsible for several types of striations in DC discharges in noble gases at low currents observed in experiments. Fundamental inconsistencies of fluid models for glow discharges have been recently highlighted [41].

The kinetic model of electrons developed in [24] and used in the present paper is well suited to describe AC discharge stratification in noble gases. According to the kinetic model, standing striations in symmetric AC discharges appear due to enhanced electron heating in low-density regions where the electric field's resistive component is maximal. We have illustrated that rapid electron diffusion in phase space enhances excitation and ionization at areas of high plasma density, supporting plasma stratification. The heat transport controlled by low-energy electrons also occurs at the rate of free electron diffusion in phase space, and the electron temperature has maximum values at the regions of high electron densities oscillating substantially in the dynamic discharge regimes. We demonstrated that electron-impact ionization plays an essential role in plasma stratification in addition to electron energy transport. In our simulations, striations did not form with the switched-off generation of electron-ion pairs in the ionization term (Eq. (14)). Switching off particle generation gradually increased the tube radius *R* and striation length in our simulations. This observation confirms that standing striations are ionization waves, not heat waves. The thermocurrent instability without ionization processes did not lead to plasma stratification in our simulations for noble gases. We have shown that the length of striations is about the radius of the tube and have observed an analog of Novak's law in AC discharges.

However, we found only one striation (S wave) type in Argon and Neon plasma simulations described in the present paper. Can other resonances be observed in AC discharges?

We have shown three characteristic time scales that control plasma stratification and discharge dynamics. The fastest is the collision time, $\tau$ and the associated momentum transfer frequency, $\nu$, which controls the isotropization of the electron distribution function. The slowest time scale is the ambipolar diffusion time, $\tau_a$. The intermediate time scale, $\tau_u$ and the associated frequency, $\nu_u$ control the electron energy relaxation rate. In our simulations, standing striations in Argon and Neon appeared at $\omega\langle\tau_a\rangle > 1$, for both the high-frequency and dynamic regimes. In the high-frequency regime, at $\omega\tau_u > 1$, the EEPF responded to the time-averaged AC electric field. In the dynamic regime, at $\nu_a < \omega < \nu_u$, the EEPF shape, electron temperature, and ionization rate oscillated substantially over the AC period. The low-frequency case, $\omega\tau_a < 1$, deserves additional studies. In this case, moving striations are expected to form at the fast electron time scale in a specific range of *E/p* and *pR*.

We have used two models to calculate the electric field in the positive column. The first model solved the Poisson equation for an applied AC voltage between the column ends. The second model separated ambipolar and resistive components for a specified discharge current. Both models calculated the radius of the tube self-consistently for a given *E/p* to balance the particle production and loss. The third model calculated the entire discharge with near-electrode sheaths and a stratified plasma column. In this model, the tube radius *R* was an input parameter, and the electric field in the stratified plasma column was calculated self-consistently for this value of *R*. The entire discharge model, including electrodes, is applicable at low driving frequencies and could be used in future work to analyze stratification in dielectric barrier discharges, which often operate in quasi-static regimes.

We used two methods for studies of plasma stratification. In our simulations with periodic boundary conditions described in Section III, the tube radius was computed for a given electric field value. Section IV prescribed the tube radius, which is closer to experimental settings. However, Section III's results are essential for studying striations' resonance properties, as described in detail in our previous paper [24] devoted to moving striations in DC discharges. The positive column of finite length acts as a cavity resonator containing a set of modes representing the S-, P-, and R-striation types. Effects of column length on the wave properties have been demonstrated in that paper for moving striations in DC discharges. However, kinetic resonances for standing striations in AC discharges have yet to be studied. The subject deserves further studies to connect the cavity resonator effects with electron bunching in spatially (*and temporally*) modulated electric fields. The time variations of plasma parameters in our simulations indicate low-frequency modulations, which depend on the striation length and the length of the positive column. We expect that the amplitude of the striations will be maximum when the length of the stratum coincides with the resonant length. Hysteresis phenomena are expected to occur when the integer number of waves inside the column is changed. These effects could be the subject of future separate studies.

In our simulations with periodic boundary conditions in Section III, we observed low-frequency modulations of plasma parameters, which depended on the applied voltage, striation length, and the length of the positive column. These modulations are sensitive to the applied voltage and indicate non-linear and cavity-resonance effects associated with the finite size of the positive column and kinetic resonances discussed above. We expect the striations' amplitude to have maximum values when the stratum length coincides with the resonant length. Hysteresis phenomena are expected to occur when the integer number of waves inside the column changes.
In our simulations with electrodes described in Section IV, there was no low-frequency modulation of plasma parameters. However, striation properties are expected to depend on the discharge gap $L$, especially when $L$ is smaller than the spatial scale of the EEPF relaxation, and the electron bunching process cannot fully develop.

We have studied the range of *E/N* where the FP kinetic equation for EEPF can be used. These are conditions when the electron mean free path, $\lambda$ is smaller than the tube radius *R*. For these conditions, we need to include metastable atoms, stepwise ionization, and Coulomb collisions for a detailed comparison with experiments. The case of $\lambda > R$ requires solving the full Boltzmann equation using PIC or grid-based methods. Future research in this direction could explore this possibility.

## VI. Conclusions

We have described a kinetic model for electrons that occupies a specific space between the fully kinetic Boltzmann treatment and hydrodynamic (fluid) models. The model, often called a Lorentz approximation, consists of the two-term spherical harmonics expansion of the electron distribution function in velocity space. The dominance of elastic scattering of electrons over inelastic collisions allows deriving a Fokker-Planck kinetic equation for EEPF in configuration-energy phase space. The numerical solution of the FP kinetic equation for electrons in $(x, u)$ phase space ($u$ is kinetic energy) was coupled to the drift-diffusion model for ions and the Poisson solver for the electric field. The FP approach is less expensive than solving the Boltzmann equation (with either grid-based or PIC methods), and it captures all the non-local effects associated with spatial *and temporal* variations of the EEPF in spatially inhomogeneous and temporally varying electric fields. We obtained standing striations in Argon and Neon AC discharges at driving frequencies exceeding the ambipolar diffusion frequency, $\omega\langle\tau_a\rangle > 1$, using a 2-level excitation-ionization model for noble gas plasma. We emphasized the need for nonlocal kinetic analysis of electrons to understand the stratification of noble gas plasmas.

We demonstrated that AC discharges in noble gases operate in a dynamic regime for a wide range of driving frequencies. In this regime, ions respond to the time-averaged electric field, whereas electrons react to the instantaneous electric field. Electron kinetics in dynamic discharges is the most sophisticated yet not fully explored. Interesting effects are associated with the disparity of time scales between ambipolar diffusion, which occurs at the ion time scale, and electron kinetics (heat transport), which occurs at the free electron diffusion time scale. Whereas the spatially non-uniform plasma density profile remains unchanged during the AC period in the dynamic discharges, complicated electron fluxes in the phase space occur due to electron heating and ionization processes controlled by instantaneous electric fields.

We have shown that standing striations in symmetric AC discharges resemble moving striations in DC discharges under similar conditions. An analog of Novak's law was observed in our simulations for standing striations. We demonstrated enhanced electron heating by the resistive electric field in the regions of low plasma density. However, the maximum electron temperature, excitation, and ionization rates occurred at the maximum plasma density due to fast electron diffusion in phase space. The most favorable conditions for stratification occur when the ambipolar and resistive field components have comparable amplitudes. Striations did not form in our simulations with the turned-off generation of electron-ion pairs. This observation confirmed thermocurrent instability without ionization processes did not lead to plasma stratification.

We have demonstrated that the FP approach is an adequate tool for self-consistent simulations of plasma stratification. Although our model describes some experimental observations, further model development is required for quantitative comparison with experiments. In future work, we plan to extend our solver to multi-dimensional problems and include metastable atoms, stepwise ionization, and Coulomb collisions to describe the effects of plasma density and transitions between different striation types. An important fundamental question is whether kinetic resonances responsible for various moving striations in DC discharges can manifest themselves in AC discharges in noble gases. The present method could be adapted to the kinetic analysis of discharge stratification in molecular gases. Experimental measurements of EEPF in stratified discharges using Langmuir probes are desirable to advance this field further.

## Acknowledgments

This work was supported by NSF project OIA-2148653 and DOE project DE-SC0021391.